\documentclass[aps,prb,twocolumn,showpacs,superscriptaddress]{revtex4}
\usepackage{amsmath,amssymb}
\usepackage{graphicx}
\usepackage{epsfig}
\usepackage{subfigure}

\newcommand{\s}{\sigma}

\renewcommand{\dag}{\dagger}

\newcommand{\nn}{\nonumber}

\newcommand{\bk}[1]{\left(#1\right)}

\newcommand{\be}{\begin{equation}}
\newcommand{\ee}{\end{equation}}
\newcommand{\bea}{\begin{eqnarray}}
\newcommand{\eea}{\end{eqnarray}}

\begin{document}
\title{Frustrated Cooper pairing and the $f$-wave supersolidity}
\author{Hsiang-Hsuan Hung}
\affiliation{Department of Physics, University of California, San Diego,
CA 92093}
\author{Wei-Cheng Lee}
\affiliation{Department of Physics, University of California, San Diego,
CA 92093}
\author{Congjun Wu}
\affiliation{Department of Physics, University of California, San Diego,
CA 92093}
\date{\today}

\begin{abstract}
Geometric frustration in quantum magnetism refers to that magnetic
interactions on different bonds cannot be simultaneously minimized.
Usual Cooper pairing systems favor  uniform spatial distributions
of pairing phases among different lattice sites without frustration.
In contrast, we propose ``frustrated Cooper pairing'' in non-bipartite 
lattices which leads to the supersolid states of Cooper
pairs.
Not only the amplitudes of the pairing order parameter but also
its signs vary from site to site.
This exotic pairing state naturally occurs in the $p$-orbital
bands in optical lattices with ultra-cold spinless fermions.
In the triangular lattice, it exhibits an unconventional
supersolid state with the $f$-wave symmetry.
\end{abstract}
\maketitle

\section{Introduction}
Frustration is one of the fundamental challenges in classic and quantum
magnetism \cite{diep2005}.
For the antiferromagnetic states in  non-bipartite lattices,
such as triangular, Kagome and pyrochlore, it is
impossible to simultaneously minimize the magnetic energy of each bond.
Consequentially, the ground state configurations are heavily degenerate.
The enhanced thermal and quantum fluctuations strongly suppress spin ordering.
The ultimate orderings often occur at much lower temperatures than
the energy scale of the antiferromagnetic coupling through
the ``order from disorder mechanism'' \cite{shender1982,henley1989}.
Furthermore, frustration provides a promising way to reach the spin liquid
states, which exhibit exotic properties including topological ordering and
fractionalization \cite{misguich2008,moessner2008}.

In the usual superfluid states of paired fermions and bosons, a uniform
distribution of the superfluid phase over the lattice sites is favored
in order to maximally facilitate phase coherence.
However, frustration can indeed occur under certain conditions.
It has been found that in the disordered superconductors near
the superconductor-insulator transition, the fluctuations of the
superfluid density can result in frustrated Josephson coupling among
superconducting grains \cite{spivak1991,zhou1998,zhou1998a}.
Recently, the striped superconductivity \cite{berg2009,berg2009a}
has been proposed for the high T$_c$ compound La$_{2-x}$Ba$_x$CuO$_4$.
The Josephson coupling between two adjacent superconducting stripes
is like in the $\pi$-junction leading to the opposite
signs of the pairing phases across the junction.
The mechanism for the frustrated coupling arises from the
interplay between superconductivity and antiferromagnetism in doped
Mott insulators.
However, an intuitive picture of the microscopic origin of this
exotic phase is still needed.

On the other hand, cold atom optical lattices have opened up a
new opportunity to investigate novel features of orbital physics
which do not exhibit in usual orbital systems of
transition metal oxides.
Bosons have been pumped into the excited $p$-orbital bands
experimentally with a long life time\cite{muller2007,wirth2010,olschlager2010}.
This metastable excited  state of bosons does not obey the ``no-node''
theory and exhibits the unconventional superfluidity with complex-valued
many-body wavefunctions breaking time-reversal symmetry spontaneously
\cite{isacsson2005,liu2006,wu2006,wu2009a}, which has already
been observed \cite{wirth2010,olschlager2010}.
For orbital fermions, large progress has been made
in the $p_{x,y}$-orbital bands in the hexagonal lattices,
whose physics is fundamentally different from that in
the $p_z$-orbital system of graphene.
The interesting physics includes the flat band structure
\cite{wu2007}, the consequential
non-perturbative strong correlation effects (e.g. Wigner crystal
\cite{wu2008b} and ferromagnetism \cite{zhang2010}),
frustrated orbital exchange interaction \cite{wu2008},
quantum anomalous Hall effect \cite{wu2008a}, and the unconventional
$f$-wave Cooper pairing \cite{lee2010}.

We are interested in bridging the above important research directions
together by introducing frustration to Cooper pairing as a new feature
of orbital physics.
In this article, we propose the ``frustrated Cooper pairing'' in the
$p_{x,y}$-band of the non-bipartite optical lattices with spinless fermions.
Due to the odd parity nature of the $p_{x,y}$-orbitals, the Josephson
coupling of the on-site Cooper pairing is frustrated.
In the strong coupling limit, the super-exchange interaction of the
pseudo-spin algebra composed of the pairing and density operators
is described by the ``antiferromagnetic'' Heisenberg model with the
Ising anisotropy.
It results in the coexistence of charge density wave and superfluidity
of Cooper pairs with a non-uniform phase pattern.
This supersolid state of Cooper pairs exhibits the $f$-wave pairing
symmetry in the triangular lattice within a large range of particle density.

Before we move on, let us explain some conceptual subtleties.
One might wonder how to justify the validity of ``frustration'' of Cooper
pairs which are usually an extended objects.
Indeed, frustration is mostly commonly defined in antiferromagnetism
of local spin moments.
However, frustration does not necessarily mean ``on-site'' physics
even in the context of antiferromagnetism in non-bipartite lattices.
For example, antiferromagnetic orders can be considered as pairing between
particles and holes in the spin triplet channel carrying nonzero momentum,
{\it i.e.}, spin density waves.
In the strong coupling limit, the particle-hole pairs are strongly bound
to be on-site, then the physics reduces to local moments
described by the Heisenberg model.
However, in the weak and intermediate coupling regimes with small
charge gaps, the systems are still locally itinerant.
The spatial extensions of the particle-hole bound states are beyond
one lattice site.
If the lattice is non-bipartite, we still have frustrated magnetism with
extended particle-hole pairs.
For example, this picture applies to the intermediate coupling regime
of the Hubbard model at half-filling in the triangular lattice.
In our case, we will consider the Cooper pairing at intermediate and strong
coupling regimes.
In the strong coupling limit, Cooper pairs are bound on a single site,
whose exchange physics can be described by the antiferromagnetic
pseudospin Heisenberg model in the charge channel.
In the intermediate coupling regime, although a Cooper pair is an
extended objects covering several lattice constants, its location
can still be defined by its center of mass.
The associated physical quantity is the pairing order parameter
at each lattice site in the mean-filed theory.
This physics can be best explained in terms of the anomalous Green's 
function $F(\vec R, \vec r; \omega)$,
where $\vec R$ is the center of mass coordinate, and $\vec r$
is the relative coordinate.
The order parameter $\Delta(\vec R)$ corresponds
to $F(\vec R, 0;\omega=0)$,
while the size of Cooper pairing is determined by the decay length
of $F(\vec R, \vec r^\prime;\omega=0)$ with respect to $r$.
In our context, frustration refers to center of mass
motion of $\Delta (\vec R)$.

This paper is organized as follows.
The model Hamiltonian and the band structure are introduced
in Sect. \ref{sect:model}.
The strong coupling analysis is
given in Sect. \ref{sect:strong}.
The mean-field theory analysis is presented in Sect.
\ref{sect:mft}, and the $f$-wave supersolid state
is presented in \ref{sect:fwave}.
Conclusions are given in Sect. \ref{sect:conclusion}.

\section{The model Hamiltonian}
\label{sect:model}

\begin{figure} [!hbt]
\centering\epsfig{file=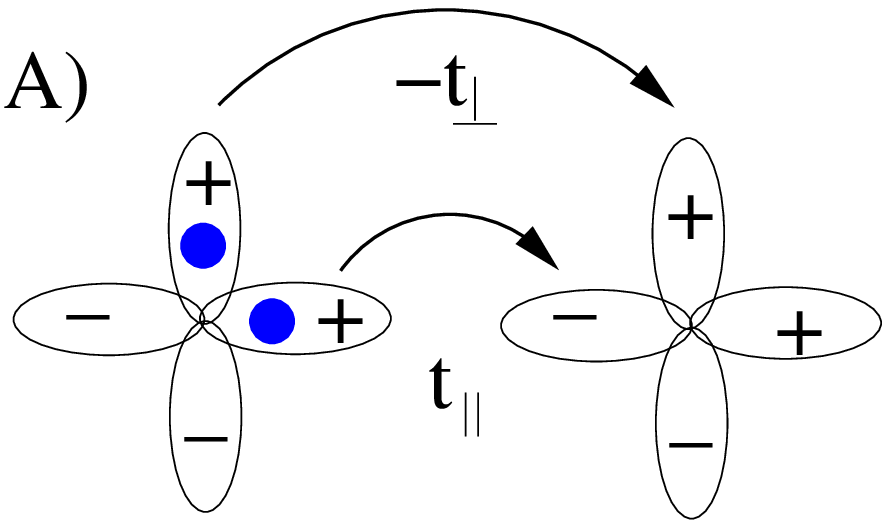,clip=1,width=0.5\linewidth,
height=0.3\linewidth,angle=0}
\hspace{2mm}
\centering\epsfig{file=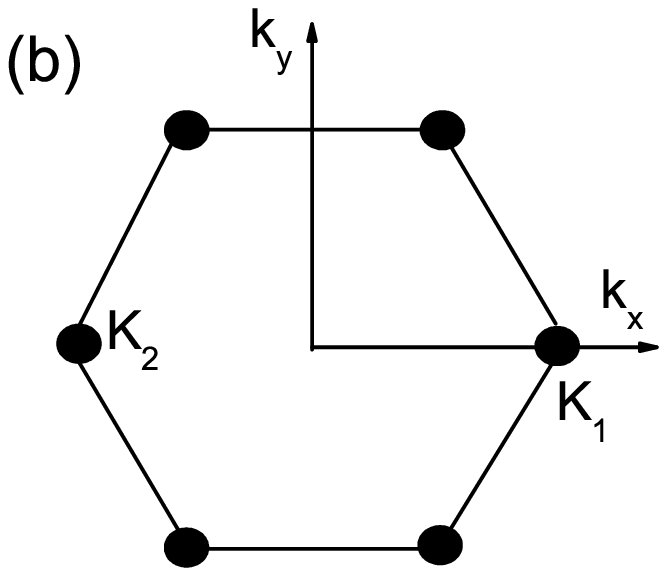,clip=1,width=0.37\linewidth,angle=0}
\caption{(a) The $\sigma$-bonding and $\pi$-bonding of the $p$-orbitals 
have opposite signs due its odd parity nature. This gives rise to the `
`antiferromagnetic''-like exchange in the change channel with 
attractive interactions as expressed in Eq. \ref{eq:exchange}.
which is frustrated in  the triangular lattice.
(b) The first Brillouin zone of the triangular lattice is
a regular hexagon. 
$K_{1,2}=(\pm \frac{4 \pi}{3a},0)$ represent two
non-equivalent vertices, and other
four are equivalent to $K_{1,2}$.} 
\label{fig:lattice}
\end{figure}

We take the 2D triangular lattice as an example, which has been
constructed experimentally by three coplanar laser beams 
\cite{becker2007}.
The optical potential on each site is approximated by a 3D anisotropic
harmonic potential with frequencies $\omega_z \gg
\omega_x=\omega_y$.
After the lowest $s$-band is fulfilled, the active orbital bands
become $p_{x,y}$.
The $p_z$-band remains empty and is neglected.
The free part of the $p_{x,y}$-orbital band Hamiltonian in
the triangular lattice filled with spinless fermions reads
\bea
\label{eq:ham0}
H_0&=&t_{\|}\sum_{\vec{r}, i=1\sim 3, \sigma}\bk{p^\dag_{L,\vec{r},i}
p_{L,\vec{r}+a\hat{e}_i,i}+h.c.}\nn \\
&-& t_\perp \sum_{\vec{r}, i=1\sim 3, \sigma}\bk{p^\dag_{T,\vec{r},i}
p_{T,\vec{r}+a\hat{e}_i,i}
+ h.c.}\nn \\
&-&\mu\sum_{\vec{r}\s}n_{\vec{r}\s}, 
\eea 
where $\vec r$ runs over all the sites; $\hat{e_1}=\hat e_x$,
$\hat e_{2,3}=-\frac{1}{2}\hat e_x \pm\frac{\sqrt 3}{2} \hat e_y$
are the three unit vectors along bond directions. 
$p_{L,i}\equiv(p_x\hat{e}_x+p_y\hat{e}_y)\cdot \hat{e}_i$ are the
longitudinal projections of the $p$-orbitals along the $\hat e_i$
direction.
More explicitly, $p_{L,1}= p_x$ and $p_{L,2(3)}=-\frac{1}{2} p_x 
\pm \frac{\sqrt 3}{2} p_y$.
The transverse projections of the $p$-orbitals along the bond
read $p_{T,i}\equiv(p_x\hat{e}_x+p_y\hat{e}_y)\cdot (\hat z
\times \hat{e}_i)$.
$n_{\vec{r}}=p_x^\dagger p_x +p_y^\dagger p_y$ is the
particle number operator; $\mu$ is the chemical potential. 
The $\sigma$-bonding $t_{\|}$ and $\pi$-bonding $t_\perp$ describe the
hoppings between $p$-orbitals along and perpendicular to the bond
direction, respectively, as depicted in Fig. \ref{fig:lattice} A.
$t_{\|}$ is positive due to the odd parity
nature of the $p$-orbitals, which is scaled 1 below.
$t_\perp$ is usually much smaller than
$t_{\|}$ because of the anisotropy of the $p$-orbitals.
The first Brillouin zone (BZ) of the triangular lattice is a
regular hexagon as plotted in Fig. \ref{fig:lattice} B.
The edge length of the first BZ is  $\frac{4 \pi}{3a}$, where $a$ is the
lattice constant.

\begin{figure} [!hbt]
\centering\epsfig{file=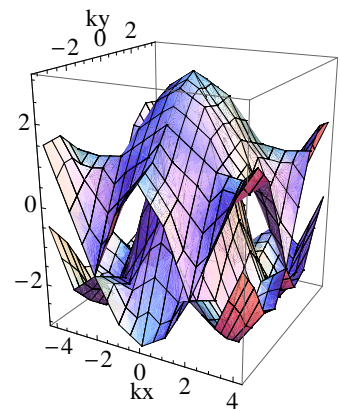,clip=1,width=0.5\linewidth,angle=0}
\caption{The Band structure of the Hamiltonian Eq. \ref{eq:ham0}.
Two bands touch at $K_{1,2}$ with the Dirac spectra, and at the 
center of the BZ with the quadratic spectra.
}
\label{fig:bandstructure}
\end{figure}

The band structure of the noninteracting Hamiltonian Eq. \ref{eq:ham0}
is displayed in Fig. \ref{fig:bandstructure} with the value of
$t_{\perp}/t_{\parallel}$ chosen as $0.2$.
The band width is around $6t_{\parallel}$. 
There is no particle-hole symmetry with respect to the zero-energy 
point, which hints an asymmetric phase diagram with respect
to half-filling for the interacting Hamiltonian introduced
in Sect. \ref{sect:strong}.
In momentum space, we define the two-component spinor 
$\psi(\vec k)=(p_x (\vec k), p_y(\vec k))^T$ 
for the $p_x$ and $p_y$-orbitals.
The Hamiltonian Eq. \ref{eq:ham0} becomes
\bea
H=\sum_{k} \psi^\dagger_\alpha(\vec k)
\Big\{ H_{\alpha\beta} (\vec k)-\mu \delta_{\alpha\beta} \Big \}
\psi_\beta(\vec k),
\eea
where the matrix kernel $H_{\alpha\beta}(\vec k)$ takes
the structure of
\bea
H(\vec k)&=&f(\vec k) +g_1(\vec k) \tau_1 + g_3(\vec k) \tau_3,
\eea
where $\tau_{1,3}$ are the Pauli matrices defined for
the basis of $p_x$, $p_y$ for the spinor of $\psi(\vec k)$;
the expressions of $f(\vec k)$, $g_1(\vec k)$ and $g_2(\vec k)$ are
\bea
f(\vec k)&=&(1-t_\perp)\sum_i \cos \vec k \cdot  \hat e_i, \nn \\ 
g_1(\vec k)&=&-\frac{\sqrt 3}{2} (1+t_\perp) (\cos \vec k \cdot \hat e_2
-\cos \vec k \cdot \hat e_3 ), \nn \\
g_2(\vec k)&=&-\frac{1}{2}(1+t_\perp) (\cos \vec k \cdot \hat e_2+
\cos \vec k \cdot \hat e_3), \nn \\
&+&(1+t_\perp) \cos \vec k \cdot \hat e_1.
\eea
The diagonalization of $H(\vec k)$ gives rise the dispersions of 
two bands as
\bea
E_\pm&=&f(\vec k) \pm (1+t_\perp)\nn \\
&\times& \sqrt{\sum_i \cos^2 \vec k \cdot \hat e_i
-\sum_{1\le a<b\le 3} \cos \vec k \cdot \hat e_a \cos \vec k \cdot 
\hat e_b }. ~~~~
\eea
These two bands touch each other at $K_{1,2}$ with the Dirac
cone-like spectra, and at the center of the BZ with the quadratic
spectra. 
When the Fermi energy is located at the Dirac points,
the other band contributes a large connected branch 
Fermi surface, thus its contributions to thermodynamic quantities 
dominate over those from the  Dirac points. 

The topology of the Fermi surfaces varies at different filling levels.
The energy minima of Eq. \ref{eq:ham0} are three-fold degenerate 
located at the middle points of the BZ edges.
The middle points of the opposite edges are equivalent up 
to a reciprocal lattice vector.
Around the band bottom, the Fermi surfaces only cut the first
band and form three disconnected elliptical pockets.
As filling increases, these pockets become connected forming
a large Fermi surface around the center of the BZ.
At the same time, the two Dirac cones contribute two 
Fermi surfaces around the $K_{1,2}$ points, which shrink
to two points when Fermi energy is right at the Dirac points.
As approaching the band top where a 
quadratic band touching exists, there are two Fermi surfaces
around the center of the BZ.

\section{Strong coupling analysis}
\label{sect:strong}

It is well-known that in the Mott-insulating state of the positive-$U$ 
Hubbard model at half-filling, its low energy physics lie in the
magnetic channel, which is captured by the antiferromagnetic
Heisenberg model\cite{auerbach1994}.
In non-bipartite lattices (e.g. the triangular lattice), the
antiferromagnetic exchange cannot be simultaneously minimized
for every bond, which leads to frustration.
Similarly, for the negative-$U$ Hubbard model, in the strong
coupling limit, the low energy physics is described by the
exchange interaction in the charge channel.
On each site, the low energy states are the doubly occupied state 
and the empty state, which can be considered the
pseudospin ``up'' and ``down'' state, respectively
\cite{micnas1990}.
The Josephson coupling between neighboring sites
plays the role of the ferromagnetic coupling in the
$xy$-direction of the pseudospin, which favors a uniform
phase distribution.
In other words, the exchange interaction in the charge
channel is unfrustrated.

This situation is fundamentally changed for the Hubbard model
of spinless fermions in $p_{x,y}$-orbitals based on the
band structure in Eq. \ref{eq:ham0}.
We add the attractive Hubbard interaction between spinless fermions 
in the $p_{x,y}$-orbital bands 
\bea
\label{eq:inter}
H_{int}&=&-U\sum_{\vec{r}} n_{\vec{r},p_x }n_{\vec{r},p_y}, 
\eea
where $U$ is positive.
The frustrated nature of Cooper pairing can be easily explained in
the strong coupling limit of $U\gg t_\parallel$. 
Similarly to the usual negative $U$ Hubbard model, we construct
the pseudospin algebra denoted
\cite{auerbach1994} as
\bea
\eta_x&=&\frac{1}{2}(p_x^\dagger p_y^\dagger+p_y p_x), \ \ \,
\eta_y=-\frac{i}{2}(p_x^\dagger p_y^\dagger-p_y p_x), \nn \\
\eta_z&=&\frac{1}{2}(n_{\vec r}-1).
\eea
Up to a normalization factor, they are the real and imaginary parts of the
pairing operator, and the particle density operator, respectively.
The low energy Hilbert space in each site consists of the doubly occupied
state and the empty state, which are eigenstate of $\eta_z$ with
eigenvalues $\pm \frac{1}{2}$, respectively.

\begin{figure}
\centering\epsfig{file=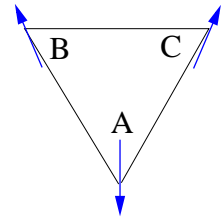,clip=1,width=0.4\linewidth,angle=0}
\caption{ The pattern of $\langle G |\vec \eta |G \rangle$ at $h=0$: 
the unit cell of three sites exhibiting the supersolid ordering.
}
\label{fig:frus-cooper}
\end{figure}

This super-exchange interaction of the pseudo-spin has a remarkable
feature that a $\pi$-phase difference is favored between pairing
order parameters $\eta_{x,y}$ on neighboring sites. As a pair hops
to the neighboring site, it gains a $\pi$-phase shift,
because the $\sigma$-bonding and $\pi$-bonding terms are with opposite signs
as depicted in Fig. \ref{fig:lattice} A.
The perturbation theory
gives rise to the anisotropic ``antiferromagnetic'' Heisenberg model
(AAFHM) in the ``external magnetic field'', 
\bea 
H_{eff}&=&\sum_{ij} J_{x,y} \big\{\eta_x (i) \eta_x (j)+\eta_y(i)\eta_y(j)\big\} 
+J_z \eta_z (i) \eta_z(j)\nn \\
&-&h\sum_i \eta_z (i),
\label{eq:exchange}
\eea
where $h=2\mu$ is the external magnetic field;
the exchange constants $J_{x,y}$ and $J_z$ read
\bea
J_{x,y}=\frac{4t_\perp t_\parallel}{U}, \ \ \
J_z=\frac{2(t_\perp^2+t_\parallel^2)}{U}.
\eea
The Ising anisotropy in Eq. \ref{eq:exchange} is because $J_z\ge J_{x,y}$.
Similar models apply to pairing problem of spinless fermions
in the $p_{x,y}$-orbital of the bipartite lattices of
square  \cite{feiguin2009} and hexagonal \cite{lee2009}  which
are not frustrated because a canonical transformation can
change $J_{x,y}$ to $-J_{x,y}$.

Eq. \ref{eq:exchange} can be interpreted as a hard core boson model
with the frustrated hopping $J_{x,y}$ and the nearest neighbor
repulsion $J_z$. It has been studied at the zero external field in
Ref. \cite{jiang2009, wang2009} which shows a supersolid
ordering\cite{andreev1969,chester1970} with a three-site unit cell as depicted
in Fig. \ref{fig:frus-cooper}. 
The competition between charge density wave and 
supersolid ordering in optical lattices has
also been studied in Refs. [\onlinecite{dao2007, koga2009}].
Site $A$ has no superfluid
component, {\it i.e.}, $\vec \eta \parallel \hat z$; sites $B$ and
$C$ develop superfluid orders with a $\pi$-phase difference.
However, the experimental realization of hard core bosons with
frustrated hopping is difficult. In comparison, our idea of the
frustrated Cooper pairing of fermions is very natural in the
$p_{x,y}$-orbital bands. Furthermore, previous studies
\cite{jiang2009,wang2009} focus on the pseudo-spin model Eq.
\ref{eq:exchange} completely neglecting the fermion degree of
freedom. In the following, instead of using  Eq. \ref{eq:exchange},
we directly study the Cooper pairing problem with the fermion
Hamiltonian in the entire filling range from $0$ to $2$.

\section{Mean-field theory at intermediate couplings}
\label{sect:mft}

\begin{figure}
\centering\epsfig{file=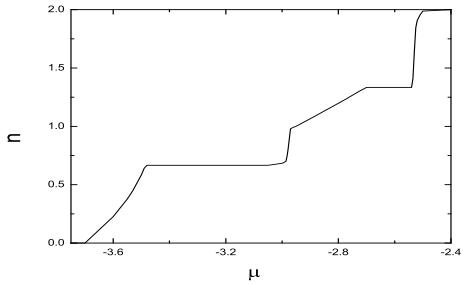,clip=1,width=0.8\linewidth, height=0.5\linewidth
,angle=0}
\caption{The average fermion number per site $n$ {\it versus}
the chemical potential $\mu$ at $U/t_\parallel=6$ and $t_\perp/t_\parallel=0.2$.}
\label{fig:mag}
\end{figure}

Below we will focus on the intermediate coupling regime and perform
the self-consistent mean-field theory to the fermionic Hubbard
model of Eq. \ref{eq:ham0} and Eq. \ref{eq:inter}.
Unlike the positive-$U$ Hubbard model with doping, in which
the mean-field theory is unreliable, in our case of negative-$U$
Hubbard model, the mean-field theory gives qualitatively
correct results for the competition between Cooper pairing
and charge density wave (CDW). 
For a detailed review, pleas refer to Ref. \onlinecite{micnas1990}.

To decouple $H_{int}$, we assume the pairing and CDW
ordering  taking an enlarge unit cell of three sites, and define
\bea
\Delta_I=\langle G|p_{\vec{r}\in I,y}p_{\vec{r}\in I,x}|G\rangle, \ \ \,
N_I=\frac{1}{2}\langle G|\hat{n}_{\vec{r}\in I}|G\rangle,
\eea
 where $I=A,B,C$ refer to the
sublattice index; $\langle G|...|G\rangle$ means the average over
the mean-field ground state. The mean-field interaction Hamiltonian
becomes: \bea H^{mf}_{int}&=&- U \sum_{\vec r, I=A,B,C}\big
\{\Delta^*_I p_{\vec{r}\in I, y}
p_{\vec{r}\in I, x} + h.c \big\}\nn \\
&+&
N_I \big \{ p^\dagger_{\vec{r}\in I, x} p_{\vec{r}\in I, x} +
p^\dagger_{\vec{r}\in I, y} p_{\vec{r}\in I, y}\big\}
\label{eq:hmf1}
\eea
Combining Eqs. \ref{eq:ham0} and \ref{eq:hmf1} and performing
Fourier transformation to momentum space,
we can obtain the resulting mean-field Hamiltonian as:
\be
H=\sum^\prime_{\vec{k}}\hat{\Psi}^\dagger(\vec{k})\left(\begin{array}{c c}
\hat{H}_s(\vec{k})&\hat{D}(\vec{k})\\
\hat{D}^\dagger(\vec{k})&-\hat{H}^*_s(-\vec{k})
\end{array}\right)\hat{\Psi}(\vec{k})
\label{mfhfinal}
\ee
where $\sum^\prime_{\vec{k}}$ means the summation only cover
half of the reduced Brillouin zone; $\hat \Psi(\vec k)$
is defined as
\bea
\hat{\Psi}(\vec{k})&=&(\phi(\vec{k})^T, \phi(-\vec{k})^\dagger),
\eea
where
$\phi(\vec{k})=[p_{A,x}(\vec{k}),p_{A,y}(\vec{k}),p_{B,x}
(\vec{k}),  p_{B,y}(\vec{k}),p_{C,x}(\vec{k}), \\ p_{C,y}(\vec{k})]$;
$H_s$ contains the free Hamiltonian Eq. \ref{eq:ham0}
combined with the CDW decoupling;
$D(\vec k)$ is the pairing part.
The order parameters are obtained self-consistently.
The above definition of order parameters are related to the
pseudospin operators through
\bea
\langle G|\eta_x(\vec{r}\in I)|G\rangle &=& {\rm Re} \Delta_I, \ \ \
\langle G|\eta_y(\vec{r}\in I)|G\rangle = {\rm Im} \Delta_I,\nn\\
\langle G|\eta_z(\vec{r}\in I)|G\rangle &=& N_I-1/2.
\label{ordertos}
\eea

Different from the ordinary BCS problem, the pairing of Eq. \ref{eq:ham0}
is not an infinitesimal instability but occurs at the finite attraction
strength.
It is because the eigen-states of the two time-reversal partners with
momentum $\vec k$ and $-\vec k$ of the free Hamiltonian Eq. \ref{eq:ham0}
have the same real polar orbital configuration.
This suppresses pairing at weak interactions because attraction
only exits in orthogonal orbitals.
With intermediate and strong interactions, pairing can occur
between different bands.
Below we present results for $t_{\perp}/t_{\parallel}=0.2$ and an
intermediate coupling and $U/t_{\parallel}=6$.
This corresponds to the effective AAFHM with the Ising anisotropy
of $J_z/J_{x,y}=2.6$.

\begin{figure}
\centering\epsfig{file=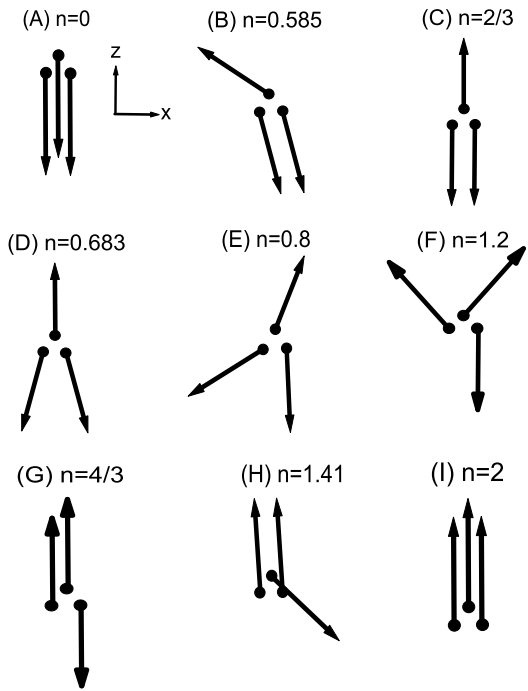,clip=1,width=0.8\linewidth, angle=0}
\caption{The real space configurations of pseudospin $\vec \eta$ on
the $xz$ plane at various fillings $n$ from (A) to (I).
A) and (I) indicate fully polarized states. (B) and (H) show three
titled vectors, where two of them have a relative $\pi$-phase to
the third one. (C) and (G) depict the CDW insulating state. (D) and (F)
exhibit an umbrella-like shape with opposite orientation. Two of them have a
$\pi$-phase difference and the third one does not own superfluid component.
(E) denotes an intermediate configuration between (D) and (F).
}
\label{fig:phase}
\end{figure}

We discuss our mean-field results in terms of the pseudospin
orientations at the three sublattices $A,B,C$. Fig. \ref{fig:mag}
shows the total fermion number per site $n=(n_A+n_B+n_C)/3$ as a
function of chemical potential $\mu$, which is the counterpart of
the magnetization in the AAFHM. The first prominent feature is the
plateaus occurring at $n=\frac{2}{3}$ and $\frac{4}{3}$, which is
corresponding to those at $\langle G|\eta_z |G\rangle
=\pm\frac{1}{3}$ observed in the study of classical ground state of
the AAFHM. These two plateaus are corresponding to CDW insulating
states without superfluidity. As shown in Fig. \ref{fig:phase}, the
corresponding pseudospin orientation for CDW insulating states is
that all the pseudospins are fully polarized along the $\hat{z}$
axis with two of sublattice along the same direction and the
remaining one along the opposite direction.

\begin{figure}
\centering\epsfig{file=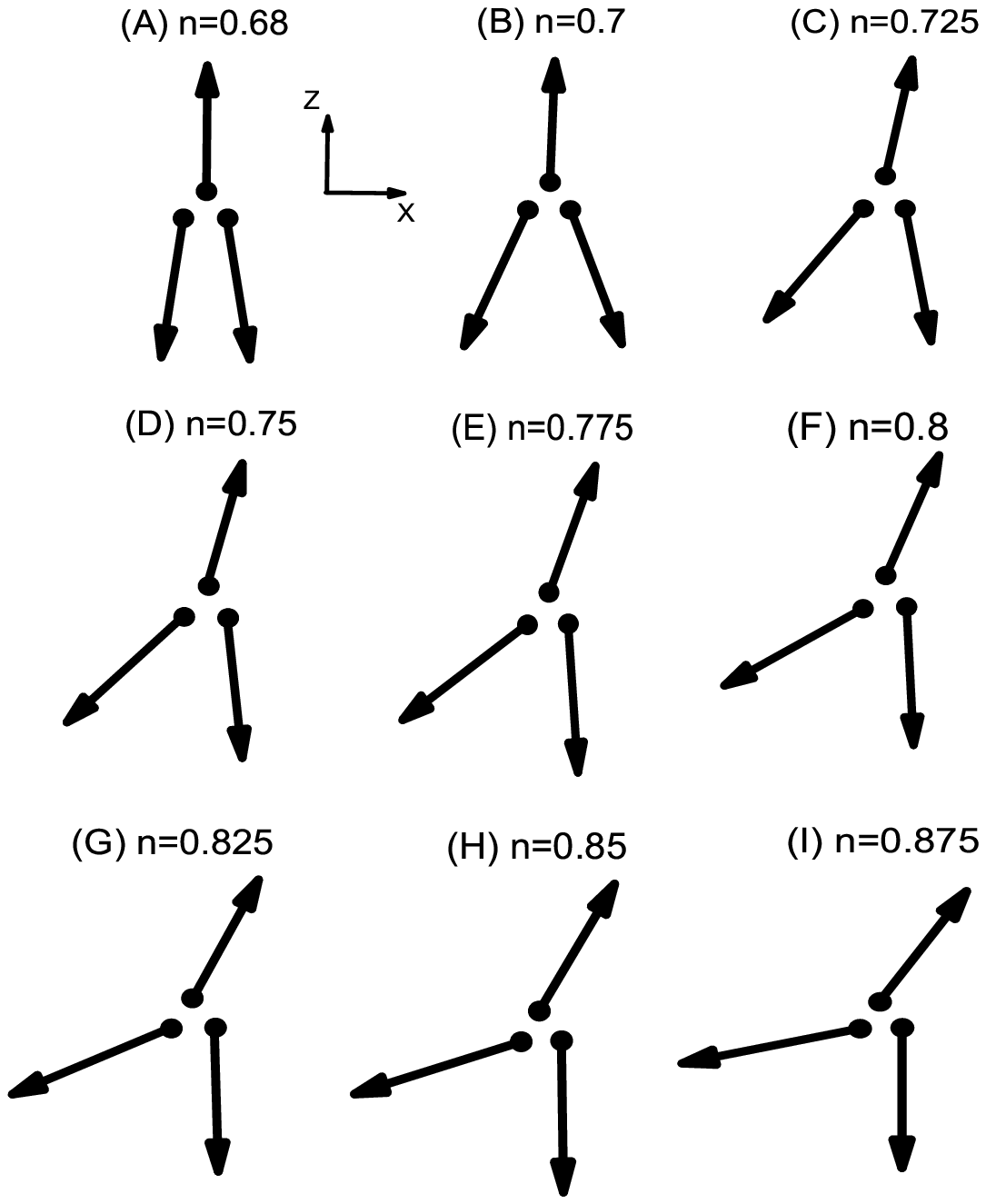,clip=1,width=0.7\linewidth, angle=0}
\centering\epsfig{file=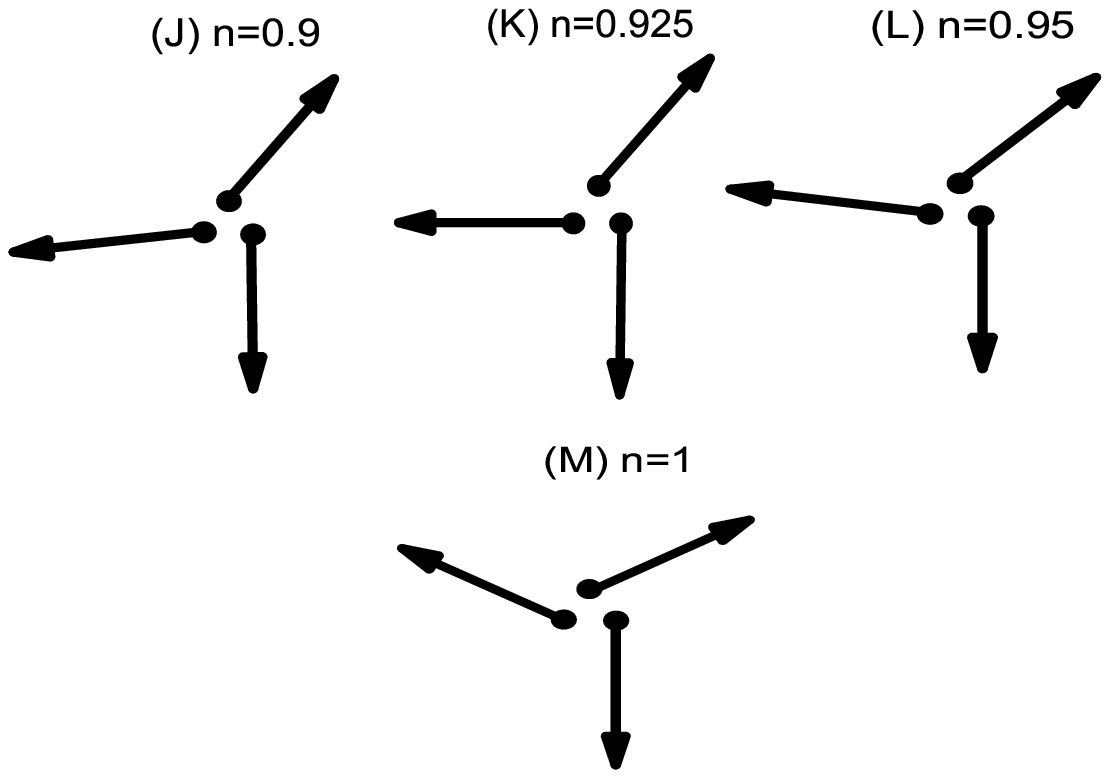,clip=1,width=0.7\linewidth, angle=0}
\caption{The smooth evolution of the two opposite ``umbrella''-like
configurations from $n=0.68$ to $n=1$.
}
\label{fig:zoomin}
\end{figure}

Although Fig. \ref{fig:mag} resembles the behaviors of the
magnetization obtained by the AAFHM \cite{miyashita1986}, two major
differences exist. First, the widths of the two CDW plateaus in Fig.
\ref{fig:mag} are different while those of the AAFHAM are the same.
We attribute this discrepancy to the different symmetry properties
between the AAFHM and the Hubbard model of Eq. \ref{eq:ham0} and Eq.
\ref{eq:inter} with the asymmetric band structure shown in Fig.
\ref{fig:bandstructure}. The AAFHM has the symmetry of the
rotation of $180^\circ$ around the $x$-axis, {\it i.e.}, $\eta_x\to
\eta_x$, $\eta_y,\eta_z\to -\eta_y,-\eta_z$ and $h\to -h$. Such an
operator corresponds to the particle-hole transformation at the
fermion level as $p_x \rightarrow i p_y^\dagger$ and $p_y^\dagger
\rightarrow i p_x$, which is not kept in the triangular lattice. As
a result, for the AAFHM, the magnetization should be an odd function
with respect to $h$ so that the lengths of the plateaus are the
same. This kind of behavior is not expected in Fig. \ref{fig:mag}.
The other difference is that at $h=0$, the ferrimagnetic state is
found in the AAFHM, and our results show the ``paramagnetic''
behavior, i.e.,  there is no jump around $n=1$. This is due to the
quantum fluctuations arising from the singly occupied states as
discussed below.

Fig. \ref{fig:phase} plots the pseudospin orientations on three
sublattices at a series of filling levels.
Except the CDW insulating states at $n=\frac{2}{3},\frac{4}{3}$,
we find that the pseudospins have non-zero
$\langle G|\eta_x|G \rangle $ and $\langle G| \eta_z| G\rangle$ in
the most part of the phase diagram, indicating the frustrated
supersolid states with non-uniform Cooper pairing density and phase.
Moreover, the phase diagram can be well-understood by the rotations of
pseudospin orientation under the magnetic field $h=2\mu$.
At $n=0$, $h$ is large along the $-\hat{z}$ direction so that all
the pseudospins are completely polarized.
As $n$ increases, the magnitude of $h$ decreases so that the
pseudospins gradually rotate upward with one of the pseudospins
($\vec{\eta}(A)$) having much faster rotating rate.
They become polarized along the $z$-direction as arriving at the
CDW insulating state $n=\frac{2}{3}$ with one pointing up and 
the other two pointing down.
The magnitudes of $\eta_{A,B,C}$ are smaller than $\frac{1}{2}$
due to quantum fluctuations.
As $n$ increases further, since $\eta^z(A)$ can not increase
anymore, $\vec{\eta}(B)$ and $\vec{\eta}(C)$ gradually turn upward
leaving $\vec{\eta}_A$ unchanged forming an umbrella configuration.
Such a state is a coexistence of superfluidity and CDW, thus
is a supersolid state. 

After a critical value $n_c\sim 0.7$, all the pseudospins start to
rotate simultaneously and continuously evolve between the two umbrella 
configurations with opposite orientations
depicted in (D) and (F) in Fig. \ref{fig:phase}, respectively.
A detailed process of evolution is plotted in Fig. \ref{fig:zoomin}
from $n=0.68$ to $n=1$.
This continuous evolution of the ground state is not present in the
AAFHM since its ground state is ferrimagnetic with non-zero
magnetization at zero field, which corresponds to $n\neq 1$
in our model.
This deviation is because the AAFHAM model is only justified
at the strong coupling limit.
The larger kinetic energy in this region leads to the less stringent assumption
of the strong-coupling.
Consequently, the quantum fluctuations arising from the singly occupied
states are enhanced, which are in disfavor of CDW but in favor of
uniform superfluidity.
The continuous evolution also explains why there is no jump at $n=1$
in Fig. \ref{fig:mag}.
Finally, the rest part of the phase diagram can be easily understood
by rotating all the pseudospins upward, and eventually all the
pseudospins are fully polarized along $+\hat{z}$ direction at $n=2$.

\section{The $f$-wave supersolid state}
\label{sect:fwave}

\begin{figure}
\centering\epsfig{file=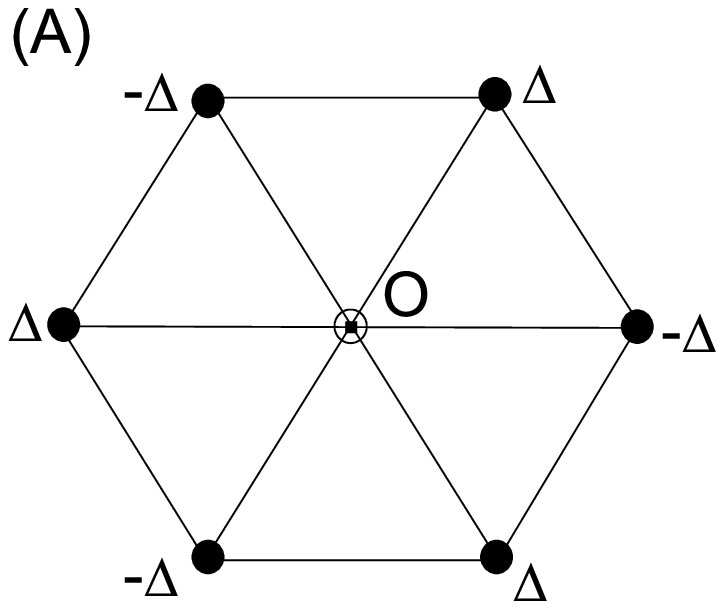,clip=1,width=0.5\linewidth,
height=0.505\linewidth, angle=0}
\centering\epsfig{file=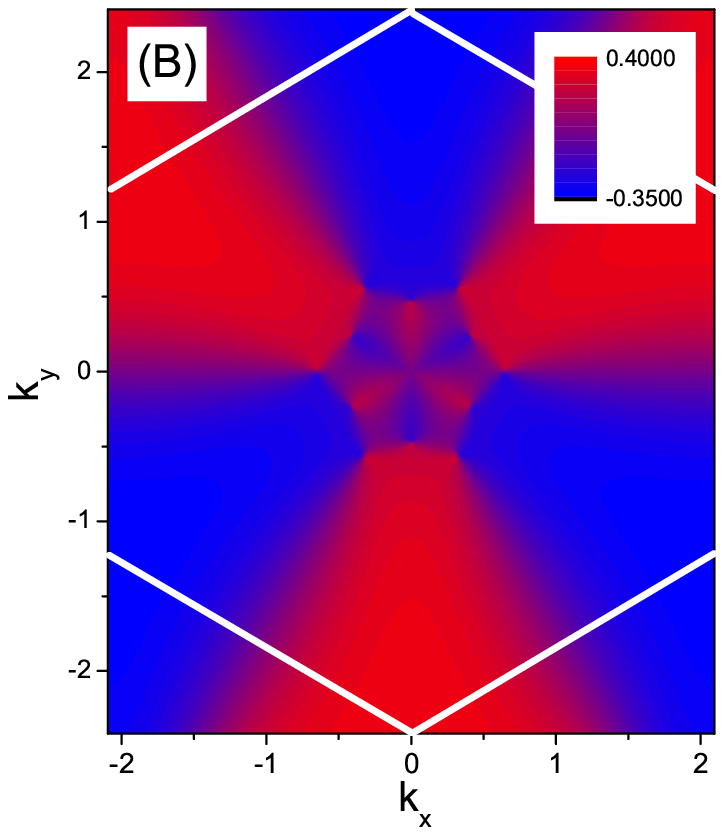,clip=1,width=0.45\linewidth,angle=0}
\caption{The $f$-wave pairing pattern in (A) real space and (B) momentum
space. The rotation of $60^\circ$ around the center site $O$ in A)
(denoted by the hollow circle) in real space or  around the center of the
reduced BZ in momentum space is equivalent to reverse the sign of the pairing
order parameters.
}
\label{fig:fwave}
\end{figure}

One remarkable feature of the frustrated Cooper pairing that we
are studying is that it can give rises to an unconventional type
supersolid state exhibiting non-$s$-wave symmetry.
For example, in Fig. \ref{fig:phase}  for a wide
region of n ($0.67\leq n\leq 0.7$, $1\leq n\leq 1.3$),
we find $\langle G|\eta_x(A) |G\rangle =0$ and
$\langle G| \eta_x(B) |G\rangle =-\langle G| \eta_x(C) |G\rangle =\Delta$.
As shown in Fig. \ref{fig:fwave} A, the signs of the pairing order
parameter are opposite in sublattices $A$ and $B$.
As a result, a spatial rotation around a site of sublattice $C$ 
at $60^\circ$ corresponds to flipping the sign of the order parameters,
which indicates the $f$-wave pairing symmetry.

The $f$-wave pairing symmetry is also manifest in the gap function 
structure in momentum space.
We calculate the intra-band pairing functions $\Delta_{nn}$ in the 
momentum space by projecting the pairing potential in Eq. \ref{eq:hmf1} 
to the band eigen-basis as:
\bea
\sum^\prime_{\vec{k}}\sum_{m,n=1}^6\,\Delta^*_{nm}(\vec{k})
\psi_n(\vec{k})\psi_m(-\vec{k})+h.c,
\eea
where
\bea
\Delta_{nm}(\vec{k}) = \big[\hat{U}^\dagger(\vec{k})
D(\vec{k})\hat{U}^*(-\vec{k})\big]_{nm},
\eea
$\hat{U}(\vec{k})$ is the unitary matrix such that
\bea
\hat{U}^\dagger(\vec{k}) H_s(\vec{k}) \hat{U}(\vec{k})
=\mbox{diag} [E_1(\vec{k}),...,E_6(\vec{k})],
\eea 
and $H_s(\vec{k})$, $D(\vec{k})$ are given in Eq. \ref{mfhfinal}.
We have confirmed that all six intra-band pairing functions have three
nodal lines and sign changes under $60^\circ$ rotation.

\begin{figure}
\centering\epsfig{file=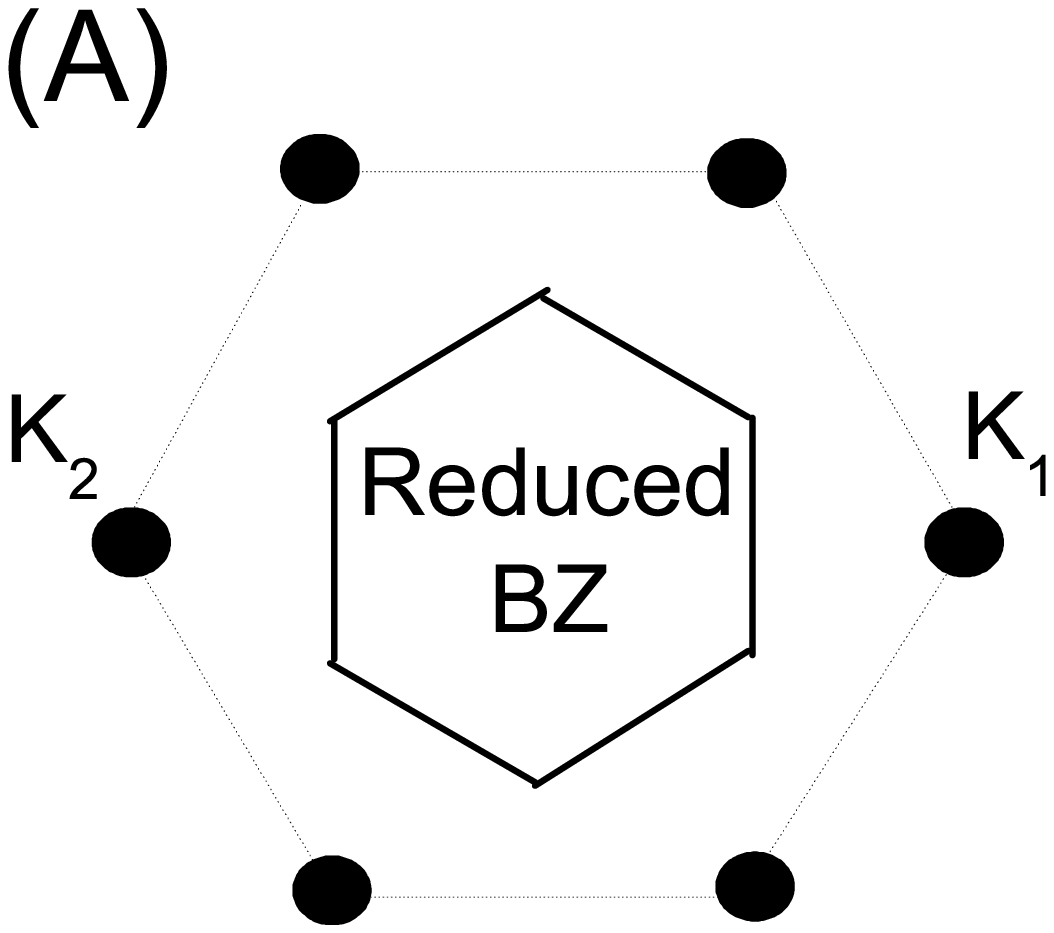,clip=1,width=0.4\linewidth, 
height=0.38\linewidth, angle=0}
\hspace{3mm}
\centering\epsfig{file=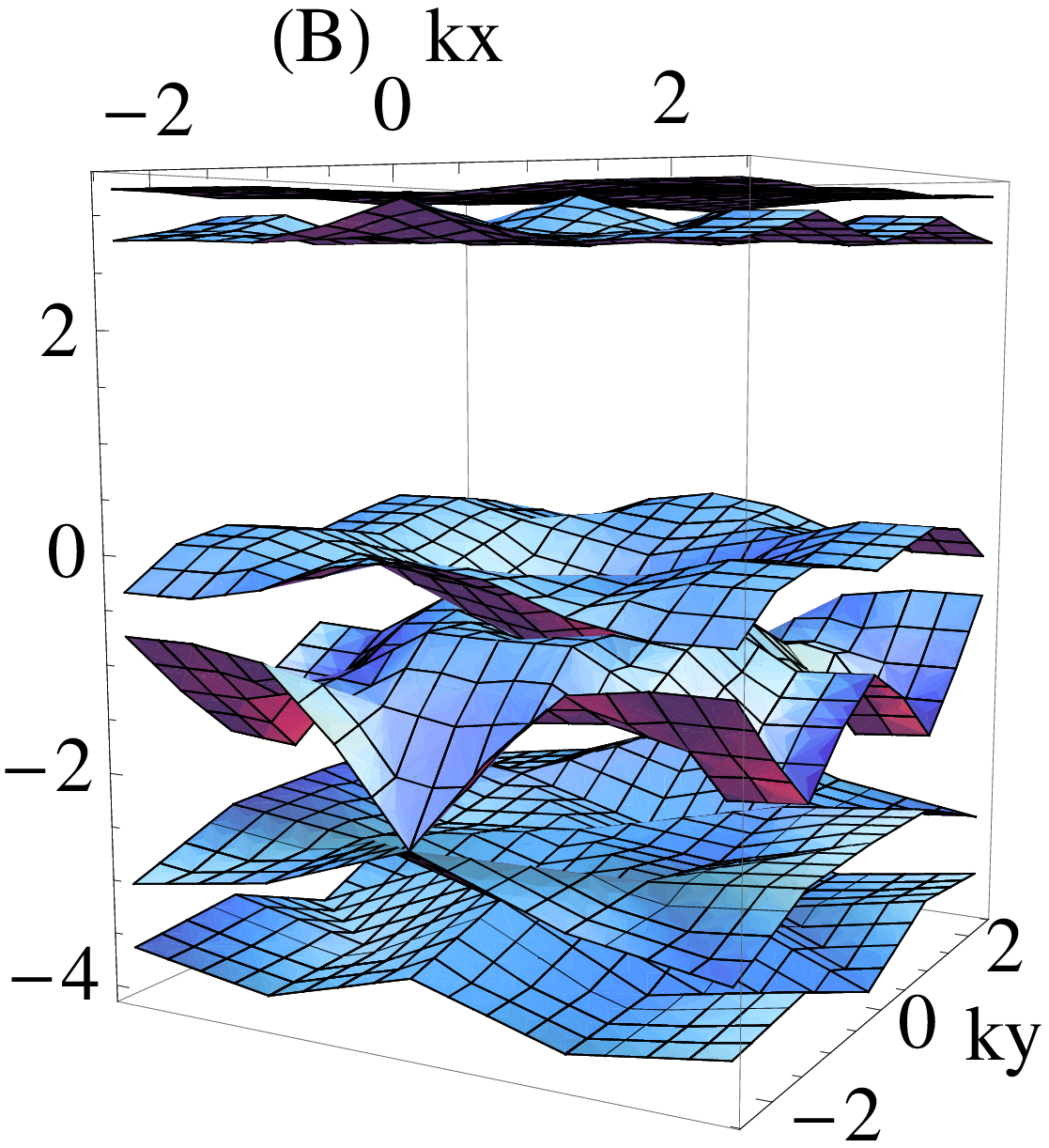,clip=1,width=0.5\linewidth,
height=0.55\linewidth,angle=0}
\caption{A) The reduced Brillouin zone (RBZ) associated with the 
enlarged three-site unit cell compared with the original BZ.
The six vertices of the RBZ are located at the centers of the
six regular triangles composed of the center of the BZ and the
vertices of the original BZ. 
B) The six bands in the reduced BZ with the three-site CDW pattern
of $n_A=n_C=1.52$, and $n_B=0.154$ for $t_\perp=0.2$ and $U=6$.
The chemical potential $\mu$ is reset to zero.
}
\label{fig:CDW}
\end{figure}

As a specific example, we present the results of the self-consistent
mean-field theory for the filling $n=1.07$ and
other parameters $t_\perp=0.2$ and $U=6$ as before. 
The system exhibits the three-site pattern of the CDW order as
$n_A=n_C=1.52$, and $n_B=0.154$, and the pairing order
parameters in the real space are $\Delta_A=-\Delta_C=0.345$
and $\Delta_B=0$.
We plot the band structure with the above CDW order parameter
but set the gap functions zero.
The reduced BZ is only $\frac{1}{3}$ of the original BZ, and
there are six bands in total as plotted in Fig. \ref{fig:CDW}.
The chemical potential $\mu$ is reset to $0$, which lies in the gap between 
the 4th and 5th bands and has no crossing with the band spectra.
As a result, although the gap functions have node lines due to
the $f$-wave symmetry, the Bogoliubov excitations  remains 
fully gapped.
We plot the gap function of $\Delta_{44}$ in Fig. \ref{fig:fwave} B
for demonstration purpose.
The nodal lines are the three lines connecting the middle points
of the opposite edges of BZ. 
Thus this is an unconventional supersolid state of frustrated
Cooper pairing with the $f$-wave pairing symmetry.
Another interesting feature is that the gap function $\Delta_{44}$
even changes sign along the radial direction.

It would also be instructive to compare our $f$-wave pairing supersolid 
state of Cooper pairs with the Fulde-Ferrell-Larkin-Ovchinnikov (FFLO) 
state \cite{fulde1964,larkin1965}.
Both cases exhibit non-uniform distributions of pairing phase in 
real space. 
However, the FFLO state completely breaks rotational symmetry.
Its pairing pattern does not form a well-defined representation of 
the lattice point group in momentum space.
In our case, it has a well-defined $f$-wave symmetry.

We also consider the extreme anisotropy limit of the vanishing
$\pi$-bonding strength, i.e., $t_\perp=0$.
The bond superexchange only results in the $J_z$-term
at the second order perturbation level in Eq. \ref{eq:exchange}.
The leading order of the hopping of the Cooper pairs occurs
through the three-site ring exchange
\bea
\Delta H= -\sum _{ijk} J^\prime [\eta_x(i) \eta_x(j)
+\eta_y(i) \eta_y (j)] \eta_z (k)
\eea
where $J^\prime=\frac{9}{2} \frac{t_\parallel^3}{U^2}$.
The hopping is frustrated for a plaquette with only one site occupied,
but it is unfrustrated for a plaquette with two sites occupied.
This means that at low fillings the phase diagram does not
change much from the case of nonzero $t_\perp$, while the system
finally evolves to a uniform pairing phase at $n$ close 2.
A more detailed analysis will be presented in a later publication.

\section{Conclusion}
\label{sect:conclusion}
In summary, we introduce the concept of ``frustrated Cooper pairing''
of spinless fermions in the $p$-orbital band in optical lattices.
The frustration occurs naturally from the odd parity of the
$p$-orbitals and is a new feature of orbital physics.
Exotic supersolid states of Cooper pairs with nonuniform
distributions of pair density and phase are obtained with an
unconventional $f$-wave symmetry.
This opens up a new opportunity to study the physics of frustrated magnet
by using the pseudo-spin algebra of the charge and pair degrees of freedom
of Cooper pairs.
This idea can also be applied to other even
more frustrated lattices, such as Kagome and pyrochlore.
In considering the  possibility of the existence of exciting spin liquid
states therein, their counterparts in terms of
``frustrated Cooper pairs'' is another interesting direction
for further exploration.

\acknowledgements
C. W. thanks J. Hirsch  for helpful discussions.
C. W., H. H. H, and W. C. L  are supported by NSF-DMR-0804775, and
AFOSR YIP program.

{\it Note added}
Upon the completion of this manuscript, we learned
the work by Cai {\it et al.} \cite{cai2009} in which a similar problem
in the square lattice is investigated.


\end{document}